\documentclass[a4paper, 12pt]{article}
\usepackage{amsmath,amssymb}
\usepackage{graphicx}

\begin{document}
\begin{titlepage}
 \renewcommand{\thefootnote}{\fnsymbol{footnote}}
$\mbox{ }$

\begin{flushright}
\begin{tabular}{l}
KEK-TH-932\\
Dec. 2003
\end{tabular}
\end{flushright}

~~\\
~~\\
~~\\

\vspace*{0cm}
    \begin{Large}
       \vspace{2cm}
       \begin{center}
         {Stability of fuzzy $S^2 \times S^2$ geometry in IIB matrix model}
\\
       \end{center}
    \end{Large}

  \vspace{1cm}

\begin{center}
           Takaaki I{\sc mai}\footnote
           {
e-mail address : imaitakaaki@yahoo.co.jp} {\sc and}
           Yastoshi T{\sc akayama}\footnote
           {
e-mail address : takaya@post.kek.jp}\\
                {\it Department of Particle and Nuclear Physics,}\\
                {\it The Graduate University for Advanced Studies,}\\
									 {\it Tsukuba, Ibaraki 305-0801, Japan}\\
\end{center}
\vfill
\begin{abstract}
\noindent
We continue our study of the IIB matrix model on fuzzy $S^2 \times S^2$.
Especially in this paper we focus on the case where the size of one of $S^2\times S^2$ is different from the other.
By the power counting and SUSY cancellation arguments, we can identify the 't Hooft coupling and large $N$ scaling behavior of the effective action to all orders.
We conclude that the most symmetric $S^2 \times S^2$ configuration where the both $S^2$s are of the same size is favored at the two loop level.
In addition we develop a new approach to evaluate the amplitudes on fuzzy $S^2 \times S^2$.
\end{abstract}
\vfill
\end{titlepage}
\vfil\eject

\section{Introduction}
\setcounter{equation}{0}
We hope the string theories tell us the dimensionality of spacetime.
Unfortunately, perturbative analysis of them suggests that the spacetime dimension should be ten rather than four.
The question is  how to derive four dimensional spacetime from string theories.
It will most likely make us understand superstring/M-theory in non-perturbative ways. 
\par
Matrix models are strong candidates for the non-perturbative formulation of superstring/M-theory\cite{BFSS}\cite{IKKT}.
Through them string theories relate to the idea of quantum spacetime
namely non-commutative geometry\cite{CDS}\cite{SW}.
In string theory, non-commutative gauge theories on flat space are realized with constant $B_{\mu \nu}$ \cite{SW} and theories on the curved space may appear with non-constant $B_{\mu \nu}$ field\cite{Myers}\cite{Alekseev}.
Non-commutative gauge theories are also obtained from matrix models with non-commutative backgrounds\cite{AIIKKT}\cite{Li}.
The gauge invariant observables of non-commutative gauge theories, the  Wilson lines were constructed through matrix models \cite{IIKK}\cite{Gross}.
They play crucial roles to elucidate the gravitational aspects of non-commutative gauge theories\cite{MRS}\cite{DK}.
\par
It is worthwhile to investigate theories on curved spaces as well as non-commutative spaces, since we live on a curved space and experimentally we have known our universe is de-Sitter space.
Particularly, $S^2$ is very interesting because it is the simplest homogeneous space and the Wick rotation of it becomes two dimensional de-Sitter space.
The homogeneous spaces  $S^2$ and $S^2 \times S^2$ are constructed at classical level by deforming the IIB matrix model\cite{MtxHom}.
In previous works we have investigated quantum corrections of these models\cite{Imai:2003vr} \cite{Imai:2003jb}.
From the results of these works we expect that the IIB matrix model single out four dimensional spacetime of spacetime among homogeneous spaces.
\par
In the context of the IIB matrix model, the four dimensionality of spacetime has been studied in several different ways: branched polymer picture\cite{BP}, complex phase effects\cite{ANG} and mean-field  approximations\cite{NS}\cite{Kyoto}.
These studies seem to suggest that the IIB matrix model predicts four dimensional spacetime.
\par
But for now it is not easy to analyze dynamics of the IIB matrix model, such as, selection of spacetime dimension and gauge groups and matter contents.
So we would like to approach general features of it through studies of concrete examples.
We think the matrix model on homogeneous spaces is a key to access these features.
In fact we have recognized that the supersymmetry of the IIB matrix model plays a key role in selecting four dimensional spacetime from our recent work\cite{Imai:2003jb}.
\par
In this paper, we investigate the IIB matrix model on the simplest four-dimensional fuzzy homogeneous background, that is, fuzzy $S^2 \times S^2$ background.
In particular we focus on the case in which the size of one of $S^2 \times S^2$ is different from that of the other $S^2$, while in the previous paper\cite{Imai:2003jb} we have focused on the case in which both are of the same size.
We compute the effective action up to the two loop level.
We can identify the 't Hooft coupling and large $N$ scaling behavior of the effective action to all order.
We argue that the most symmetric $S^2 \times S^2$ configuration where the both $S^2$ are of the same size minimizes the effective action up to two loop level.
\par
We stress that loop corrections to the effective action play an important role.
The tree level effective action has no local minimum (in terms of 't Hooft coupling $\lambda$, see eq. (\ref{tree})).
This means that the fuzzy $S^2 \times S^2$ configuration is not stable at the tree level.
Actually we find that two loop corrections allow the (quantum) effective action to have a local minimum, so that the situation is drastically changed.
In addition we develop a new approach to evaluate the amplitudes on fuzzy $S^2 \times S^2$.
\par
This paper is organized as follows. 
In section 2, we review fuzzy $S^2$ and the IIB matrix model on fuzzy $S^2 \times S^2$ background.
In section 3, we investigate the effective action of the IIB matrix model.
In addition we develop a new approach to evaluate the amplitudes on fuzzy $S^2\times S^2$.
We conclude in section 4 with discussions.
\section{IIB matrix model on fuzzy $S^2 \times S^2$ background}
Let us briefly review the fuzzy $S^2$.
$S^2$ is a homogeneous space since $S^2=SU(2)/U(1)$.
In $SU(2)$ group, there are three generators $j_1, j_2, j_3$ which satisfy the commutation relation of the angular momenta:
\begin{eqnarray}
[j_1, j_2]=i j_3.
\end{eqnarray}
This commutation relation can be realized with finite matrices since such algebra is compact.
One can obtain a non-commutative space by identifying $j_i(i=1,2,3)$ with the coordinates $x_i(i=1,2,3)$.
\par
Let us adopt $j_i$ as $N=2l+1$ dimensional representation of spin $l$.
The Casimir operator 
\begin{eqnarray}
\sum_{i=1}^{3} {x_i}^2 = \sum_{i=1}^{3} {j_i}^2 = l(l+1) \label{Casimir}
\end{eqnarray}
gives the size of the $S^2$, strictly speaking, the square radius of the sphere.
It is easy to extend this construction to fuzzy $S^2 \times S^2$, which we will use as a background.
\par
In the remaining part of this section we summarize the IIB matrix model on a fuzzy $S^2 \times S^2$ background \cite{Imai:2003jb}.
The IIB matrix model action \cite{IKKT} is
\begin{eqnarray}
S_{IIB} = -{1 \over 4} \sum_{\mu, \nu=0}^{9} Tr [A_\mu, A_\nu]^2
	+{1\over 2}\sum_{\mu=0}^{9}Tr\bar{\psi}\Gamma^\mu [A_\mu, \psi],\label{IIBaction}
\end{eqnarray}
where $A_\mu$ and $\psi$ are $N \times N$ matrices.
Since this model has the translational invariance
\begin{eqnarray}
A_\mu \to A_\mu +c_\mu
\end{eqnarray}
and
\begin{eqnarray}
\psi \to \psi + \epsilon,
\end{eqnarray}
we remove these zero modes by restricting $A_\mu$ and $\psi$ to be traceless.
\par
Let us split the field $A_\mu$ into the fuzzy $S^2 \times S^2$ background and a fluctuating quantum field:
\begin{eqnarray}
A_\mu &\to& p_\mu + a_\mu.
\end{eqnarray}
Here we take the background $p_\mu$ as follows:
\begin{eqnarray}
\begin{array}{cc}
p_\mu = f (j_\mu \otimes 1) \otimes 1_n,  & (\mu=8, 9, 0)\\
p_\mu = f (1 \otimes {\tilde{j}}_\mu) \otimes 1_n, & (\mu=1, 2, 3)\\
p_\mu=0,  & (others) 
\end{array} \label{DefP}
\end{eqnarray}
where $j_\mu$ and ${\tilde{j}}_\mu$ are angular momentum generators of spin $l_1$ and $l_2$ respectively.
Here we have introduced a free parameter $f$, which assigns the overall size of the background (see (\ref{radius})).
These $p_\mu$ satisfy the Lie algebra of $SU(2) \times SU(2)$:
\begin{eqnarray}
\left[p_\mu, p_\nu \right] &=& i f_{\mu \nu \rho} p_\rho, \\
f_{\mu \nu \rho} &=&
	\left\{
	\begin{array}{cc}
		f \epsilon_{\mu \nu \rho} & \quad (\mu, \nu, \rho) \in (8, 9, 0)\\
		f \epsilon_{\mu \nu \rho} & \quad (\mu, \nu, \rho) \in (1, 2, 3)\\
		0 & (others).
	\end{array}
	\right.
\end{eqnarray}
Casimir operators of this algebra
\begin{eqnarray}
\sum_{\mu=8,9,0} (p_\mu)^2=f^2 l_1(l_1+1), \quad
\sum_{\mu=1,2,3} (p_\mu)^2=f^2 l_2(l_2+1) \label{radius}
\end{eqnarray}
determine the size of each $S^2$ of fuzzy $S^2 \times S^2$ such as (\ref{Casimir}).
Since $p_\mu$ acts on the $n$ copies of spin $l_1$ and spin $l_2$ representations respectively(see (\ref{DefP})), the size of matrices $A_\mu$ and $\psi$ is $N=n(2l_1+1)(2l_2+1)$.
This background represents $n$ coincident fuzzy $S^2 \times S^2$, and so that this theory is a $U(n)$ gauge theory on the fuzzy $S^2 \times S^2$ space.
In the $l_2=0$ limit this background reduces to a 2 dimensional space $S^2$.
In this paper, it is always assumed that $l_1$ and $l_2$ are large, which means $n \ll N$.
\par
In this background the action (\ref{IIBaction}) becomes
\begin{eqnarray}
S&=& S_{IIB} + S_{gst} + S_{gf} \nonumber\\
	&=& -{1 \over 4} Tr \left[p_\mu, p_\nu \right]^2
		+ {1 \over 2} Tr {\overline \psi} \Gamma^\mu \left[p_\mu, \psi \right]
		+ Tr{b \left[p_\mu, \left[p_\mu, c \right] \right]} \nonumber\\
	& &-Tr a_\rho \left( \left[p_\mu, \left[ p_\rho, p_\mu \right]\right] \right)
		+ {1 \over 2} Tr {\overline \psi} \Gamma^\mu \left[a_\mu, \psi \right]
		+Tr{b \left[p_\mu, \left[a_\mu, c \right] \right]}, \nonumber\\
	& &+{1 \over 2} Tr[a_\mu(\delta_{\mu \nu} P^2 + 2if_{\mu \nu \rho}P_\rho)a_\nu] \nonumber\\
	& & -Tr \left[p_\mu, a_\nu \right] \left[a_\mu, a_\nu \right] \nonumber\\
	& &-{1 \over 4} Tr{\left[a_\mu, a_\nu \right]^2}.  \label{IIBbg}
\end{eqnarray}
where we have added a ghost term $S_{gst}=Tr{b \left[p_\mu, \left[p_\mu+a_\mu, c \right] \right]}$
and a gauge fixing term $S_{gf}=-{1\over2}Tr[p_\mu, a_\mu][p_\nu, a_\nu]$
and introduced the adjoint operator $P_\mu X= [p_\mu, X]$.
\section{Effective action}
In this section we study fuzzy $S^2\times S^2$ geometry by investigating the effective action of the model (\ref{IIBbg}).
In particular, we investigate which configuration minimizes it in terms of ratio of the radii of $S^2 \times S^2$.
\subsection{Tree \& one loop effective action}\label{sec:tree1loop}
The tree level effective action is the first term in RHS of (\ref{IIBbg}) :
\begin{eqnarray}
\Gamma_{tree}&=&-{1 \over 4} Tr [p_\mu, p_\nu]^2\nonumber\\
	&=& {f^4 \over 2} N \left(l_1(l_1+1)+l_2(l_2+1)\right)\nonumber\\
	&\simeq& {(2\pi)^2 \over \lambda^2}\left({r+{1 \over r} \over 2}\right)nN \label{tree}
\end{eqnarray}
In the third line we have used $l_1, l_2\gg1$ and $r\equiv l_2/l_1$.
Here we have also used  't Hooft coupling $\lambda^2=(4 \pi)^2 n^2/(f^4 N)$, which will be investigated later.
It is noted that $r$ is, roughly speaking, the ratio of the radii of two $S^2$s. 
It is because, for large $l_1$ and $l_2$, the ratio of the radii becomes
\begin{eqnarray}
\sqrt{{\sum_{\mu=1,2,3} (p_\mu)^2 \over \sum_{\mu=8,9,0} (p_\mu)^2}}
	=\sqrt{l_2(l_2+1) \over l_1(l_1+1)}
	\simeq {l_2 \over l_1},
\end{eqnarray}
which is $r$.
In the previous paper\cite{Imai:2003jb} we have investigated $r=0, 1$ cases which have been called 2d and 4d limit respectively.
Now we are interested in $r \neq 0, 1$ case.
The $r$-dependence of the tree level effective action is quite natural because the effective action must have a symmetry associated with the exchange of one of $S^2 \times S^2$ with the other, that is, with the exchange of $l_1$ with $l_2$.
It means that the $r$-dependence of it, which is to say, $(r+{1 \over r})=({l_2 \over l_1}+{l_1 \over l_2})$, is consistent with this argument.
\par
It is noticed that we have three parameters to minimize the effective action: one is $f$, another is $n$ of the gauge group $U(n)$,
the other is $r$.
At the tree level the effective action is minimized at $n=1$ and $r=1$, that is, the two spheres are the same size if we fix $\lambda$.
Since it is not stationary with respect to $f$ unless it vanishes, we have to investigate higher order corrections to obtain physical predictions.
\par
The leading terms of the one loop effective action in the large $N$ limit is
\begin{eqnarray}
\Gamma_{1st}&\simeq&-Tr\Big(({1\over P^2}^4)F_{\mu\nu}F_{\nu\lambda}F_{\lambda\rho}
F_{\rho\mu}\Big)
-2Tr\Big(({1\over P^2}^4)F_{\mu\nu}F_{\lambda\rho}F_{\mu\rho}
F_{\lambda\nu}\Big)\nonumber \\
&&+{1\over 2}Tr\Big(({1\over P^2}^4)F_{\mu\nu}F_{\mu\nu}F_{\lambda\rho}
F_{\lambda\rho}\Big)
+{1\over 4}Tr\Big(({1\over P^2}^4)F_{\mu\nu}F_{\lambda\rho}F_{\mu\nu}
F_{\lambda\rho}\Big)\nonumber \\
&=&3n^2Tr({1\over P^2})^2-6n^2 Tr{({P_1}^2)^2+({P_2}^2)^2\over (P^2)^4}
+n^2Tr({1\over P^2})^3\nonumber \\
&=&3n^2\sum_{j=0}^{2l_1} \sum_{p=0}^{2l_2} {(2j+1)(2p+1)\over 
(j(j+1)+p(p+1))^2}-
6n^2\sum_{j=0}^{2l_1} \sum_{p=0}^{2l_2} {(2j+1)(2p+1)(j^2(j+1)^2+p^2(p+1)^2)\over 
(j(j+1)+p(p+1))^4}\nonumber \\
&&+n^2\sum_{j=0}^{2l_1} \sum_{p=0}^{2l_2}{(2j+1)(2p+1)\over 
(j(j+1)+p(p+1))^3},
\end{eqnarray}
where we have defined $F_{\mu\nu}X=[f_{\mu\nu},X]$ and $f_{\mu\nu}=i[p_\mu, p_\nu]$.
This can be estimated as
\begin{eqnarray}
\Gamma_{1st} \simeq-n^2 \ln \left[{N \over n} \left( {r + {1 \over r} \over 2}\right)^{-1} \right] \label{1LpEffAction}
\end{eqnarray}
for large $l_1, l_2$ but fixed $r=l_2/l_1$.
The $r$-dependence of (\ref{1LpEffAction}) is also consistent with the argument we just gave.
The one loop corrections are found to be sub-leading compared to the tree level action in the large $N$ limit.
\par
\subsection{Two loop corrections} \label{subsec:asymptotic}
\begin{figure}
 \begin{center}
  \includegraphics[height=3cm]{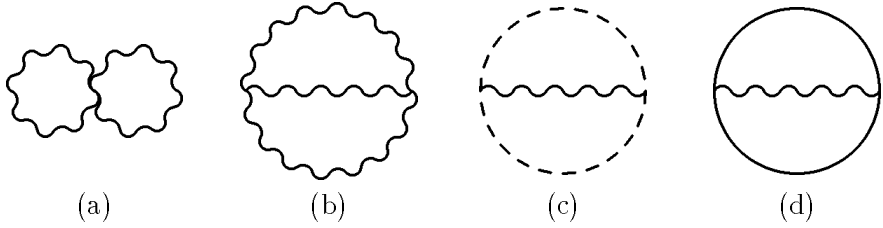}
 \end{center}
\caption	{Feynman diagrams which contribute to two loop effective action:
	(a) and (b) represents contributions form gauge fields.
	(c) involves ghosts and (d) fermions respectively.}
\label{Fig:2loopdiagrams}
\end{figure}

The diagrams that contribute to the two loop effective action are illustrated in Fig.\ref{Fig:2loopdiagrams}.
The two loop effective action is
\begin{eqnarray}
\Gamma_{2nd} \simeq n N \lambda^2 {2F(l_1, l_2) \over (4 \pi)^2} \label{Gamma2nd},
\end{eqnarray}
with\footnote{$\Gamma_{2nd}$ has been calculated in the appendix B of \cite{Imai:2003jb}, but we have found an error in (B.6). After the correction, eq. (\ref{Gamma2nd}) differs from the orignal one by factor 2. Note that this correction does not change the qualitative conclusion of \cite{Imai:2003jb}}
\begin{eqnarray}
F(l_1,l_2)&=&
2 \sum^{2l_1}_{\substack{j_1=0 \\
                     j_1=k_1 \neq 0}}
\sum^{2 l_2}_{k_1=0}
\sum^{2l_1}_{\substack{j_2=0 \\
                     j_2=k_2 \neq 0}} 
\sum^{2l_2}_{k_2=0}
\sum^{2l_1}_{\substack{j_3=0 \\
                     j_3 = k_3\neq0}} 
\sum^{2l_2}_{k_3=0}\nonumber\\
&&\frac{(2j_1+1)(2j_2+1)(2j_3+1)(2k_1+1)(2k_2+1)(2k_3+1)}
{[j_1(j_1+1)+k_1(k_1+1)][j_2(j_2+1)+k_2(k_2+1)][j_3(j_31+1)+k_3(k_3+1)]}\nonumber \\
		&&\times
		\left\{
		\begin{array}{@{\,}ccc@{\,}}
			j_1 & j_2 & j_3 \\
			l_1  & l_1 & l_1
		\end{array}
		\right\}^2
\times
		\left\{
		\begin{array}{@{\,}ccc@{\,}}
			k_1 & k_2 & k_3 \\
			l_2  & l_2 & l_2
		\end{array}
		\right\}^2.\label{F(l1,l2)}
\end{eqnarray}
where $\{\cdots\}$ is Wigner's 6j symbol\cite{Edm}.
Here we have used the 't Hooft coupling $\lambda$, which will be clarified later.
We have ignored contributions from non-planar diagrams because they are sub-leading by comparison with those of planar diagrams in the large $N$ limit.
\par
\begin{figure}
 \begin{center}
  \includegraphics[height=8cm]{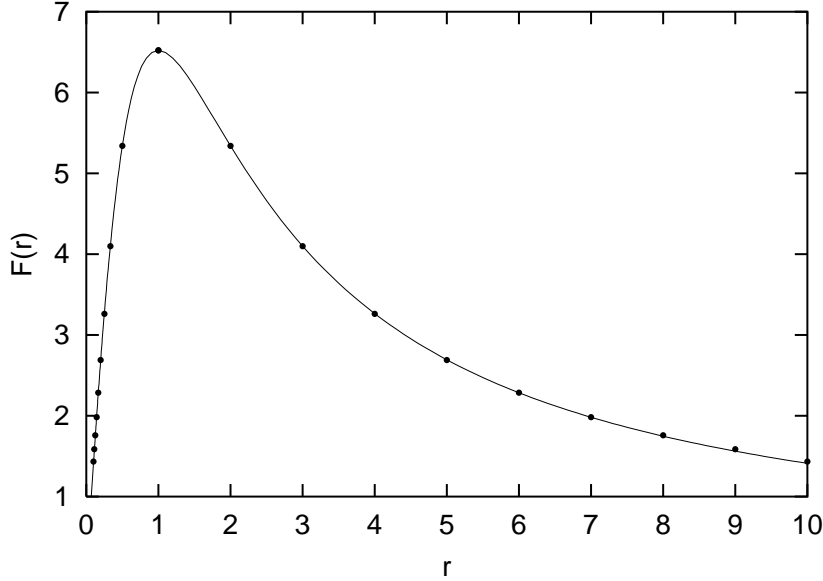}
 \end{center}
\caption{$F(r)$ against $r(=l_2/l_1)$:The solid circles are obtained by the method in the section \ref{subsec:asymptotic} and the solid line by that of the section \ref{subsec:intApprx}}
\label{Fig:Fr}
\end{figure}

For large $l_1, l_2$ while preserving $r=l_2/l_1$, we can evaluate $F(l_1, l_2)$ numerically as follows.
Let us focus for example on the case $r=2$. 
We compute $F(l_1,l_2=2l_1)$ up to $l_1=15$ (the values are represented by open circles in the fig.\ref{Fig:F(l1,2xl2)}).
\begin{figure}
 \begin{center}
  \includegraphics[height=8cm]{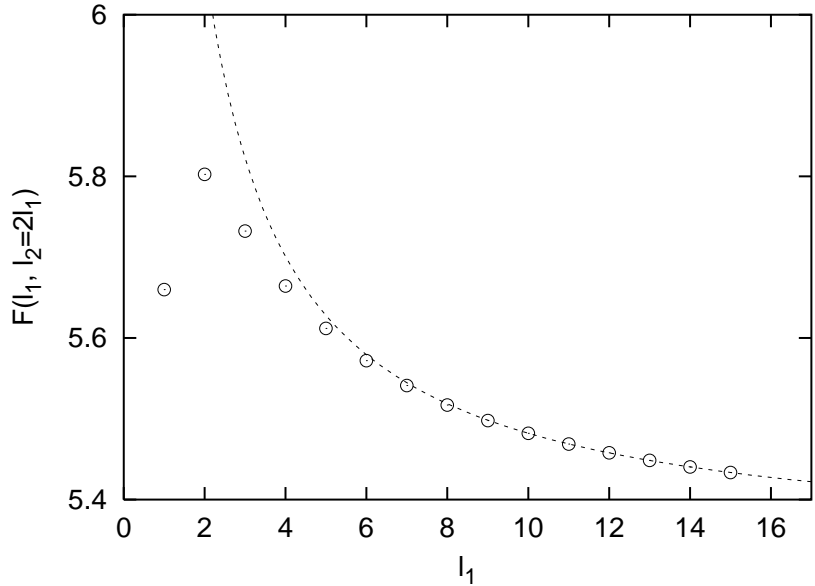}
 \end{center}
\caption{Plot of $F(l_1, l_2=2l_1)$ as a function of $l_1$ with open circles.
			The dotted line($f(l_1) \simeq 5.34+1.45/l_1$) is obtained by fitting the data greater than $l_1=10$.}
\label{Fig:F(l1,2xl2)}
\end{figure}
Fitting a function $f(l_1)=a+{b \over l_1}$ to the data of $F(l_1,2l_1)$ above $l_1=10$, we read off $a\simeq5.34$ and $b\simeq1.45$.
Therefore $F(l_1, l_2=2 l_1)\simeq5.34$ for large $l_1$.
This evaluation applies to $r \neq 2$ cases also.
In the future $F(l_1,l_2)$ is called as $F(r)$, since $F(l_1, l_2)$ with large $l_1,l_2$ but fixed $r=l_2/l_1$ depends only on $r$.
Then (\ref{Gamma2nd}) becomes
\begin{eqnarray}
 \Gamma_{2nd} \simeq nN\lambda^2{2F(r) \over (4\pi)^2} \label{2loop}
\end{eqnarray}
where $F(r)$ behaves as in the fig. \ref{Fig:Fr}.
\subsection{Another approach to evaluate two loop corrections} \label{subsec:intApprx}
In the last sub-section we obtained the two loop corrections $\Gamma_{2nd}$ by evaluating the asymptotic forms of $F(l_1, l_2=rl_1)$ for large $l_1$.
In this section we develop an alternative approach to evaluate two loop corrections.
\par
\begin{figure}
 \begin{center}
  \includegraphics[height=5cm]{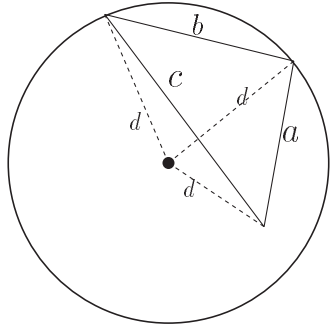}
 \end{center}
\caption{The tetrahedron whose edges are $a=j_1+{1\over2}, b=j_2+{1\over2}, c=j_3+{1\over2}$, and $d=l+{1\over2}$. This tetrahedron can be drawn like this figure, that is, its three vertices are on the sphere of radius $d$ and the other vertex is at the origin.}
\label{Fig:tetrahedron}
\end{figure}

 
The key for this new approach is the Wigner's asymptotic estimate\cite{Wigner}\cite{Ponzano}:
for large $j_1, j_2, j_3$ and $l$ the 6j symbol is related to the volume $V$ of the tetrahedron whose edges are $j_1+{1\over2}, j_2+{1\over2}, j_3+{1\over2}, l+{1\over2}, l+{1\over2}$, and $l+{1\over2}$ (see Fig.\ref{Fig:tetrahedron})
\begin{eqnarray}
		\left\{
		\begin{array}{@{\,}ccc@{\,}}
			j_1 & j_2 & j_3\\
			l   & l   & l
		\end{array}
		\right\}^2
	\simeq
{1 \over 24\pi V(j_1,j_2,j_3,l)} \label{Wigner}
\end{eqnarray}
where
\begin{eqnarray}
V^2=\frac{1}{2^3(3!)^2}
		\left|
		\begin{array}{@{\,}ccccc@{\,}}
			 0             & (l+{1\over2})^2 &(l+{1\over2})^2  &(l+{1\over2})^2  & 1\\
			(l+{1\over2})^2& 0               &(j_3+{1\over2})^2&(j_2+{1\over2})^2& 1\\
			(l+{1\over2})^2&(j_3+{1\over2})^2& 0               &(j_1+{1\over2})^2& 1\\
			(l+{1\over2})^2&(j_2+{1\over2})^2&(j_1+{1\over2})^2& 0               & 1\\
			 1             & 1               & 1               & 1               & 0
		\end{array}
		\right|
\end{eqnarray}
(In the case $V^2$ is zero or negative, in other words the tetrahedron with given edges does not exist, 6j symbols is approximated to zero.)
\par
Firstly we must check whether this approximation works or not.
A naive way to check it is to substitute Wigner's asymptotic form (\ref{Wigner}) into all of the 6j symbols in (\ref{F(l1,l2)})
\begin{eqnarray}
F_{Wig}(l_1,l_2)&=&
2	\sum^{2l_1}_{j_1,j_2,j_3=0}
	\sum^{2l_2}_{k_1,k_2,k_3=0}\nonumber\\
&&\frac{(2j_1+1)(2j_2+1)(2j_3+1)(2k_1+1)(2k_2+1)(2k_3+1)}
{[j_1(j_1+1)+k_1(k_1+1)][j_2(j_2+1)+k_2(k_2+1)][j_3(j_31+1)+k_3(k_3+1)]}\nonumber \\
&&\times W(j_1,j_2,j_3,l_1)W(k_1,k_2,k_3,l_2),\label{Fwig}
\end{eqnarray}
where $W(j_1,j_2,j_3,l_1)={1 \over 24 \pi V(j_1,j_2,j_3,l_1)}$ and similarity for $W(k_1,k_2,k_3,l_2)$.
We can compare  (\ref{F(l1,l2)}) with (\ref{Fwig}) by evaluating them numerically(See Table \ref{tbl:Wig}).
\begin{table}
\begin{center}
\begin{tabular}{|c|c|c|c|}
\hline
$l_1$&$l_2$&$F(l_1,l_2)$&$F_{Wig}(l_1,l_2)$\\ \hline
5    &   5 & 6.7216   & 6.5118 \\ \hline
10   &  10 & 6.6352   & 6.7208 \\ \hline
15   &  15 & 6.5954   & 6.6752 \\ \hline
20   &  20 & 6.5750   & 6.4954 \\ \hline
30   &  30 & 6.5551   & 6.5822 \\ \hline
\end{tabular}
\end{center}
\caption{Numerical evaluation of (\ref{F(l1,l2)}) and (\ref{Fwig}). It is clear that the both values approaches each other as we increase $l_1$ and $l_2$.}
\label{tbl:Wig}
\end{table}
The table \ref{tbl:Wig} shows that $F_{Wig}(l_1,l_2)$ comes closer to $F(l_1,l_2)$ as we increase $l_1$ and $l_2$.
It seems a bit surprising that this approximation works so well.
It will be clarified later why is the case.
We expect that $F(l_1,l_2)$ and $F_{Wig}(l_1,l_2)$ are equal in the large $l_1$ and $l_2$ limit and that this approximation is exact in this limit.
\par
Moreover in the limit $l_1$ and $l_2\to\infty$ the summations in (\ref{Fwig}) can be approximated by integrations.
As an example let us focus on the case $l_1=l_2$.
After rescaling variables ${j_1\over l_1} \to j_1$, ${k_1\over l_1} \to k_1$, and so forth and taking the limit $l_1 \to \infty$, we obtain
\footnote{$\int_0^2 dj_1 dj_2 dj_3 dk_1 dk_2 dk_3=\int_0^2 dj_1 \int_0^2 dj_2 \int_0^2 dj_3 \int_0^2 dk_1 \int_0^2 dk_2 \int_0^2 dk_3$}
\begin{eqnarray}
F=2\int_0^2 dj_1 dj_2 dj_3 dk_1 dk_2 dk_3
	{2^6 j_1 j_2 j_3 k_1 k_2 k_3 \over (j_1^2+k_1^2)(j_2^2+k_2^2)(j_3^2+k_3^2)}
\times
W(j_1,j_2,j_3)W(k_1,k_2,k_3)\label{F}
\end{eqnarray}
where the Wigner asymptotic estimate changes the 6j symbols as
\begin{eqnarray}
l_1^3
	\left\{
		\begin{array}{ccc}
		j_1 & j_2 & j_3 \\
		l_1 & l_1 & l_1
		\end{array}
	\right\}^2
\simeq
W(j_1,j_2,j_3)
\equiv
	\left\{
		\begin{array}{cc}
			{1 \over 24 \pi V(j_1,j_2,j_3)} & (V^2 > 0) \\
			0                  & (else)
	\end{array}
\right., \label{WigEst}
\end{eqnarray}
with
\begin{eqnarray}
V^2=\frac{1}{2^3(3!)^2}
		\left|
		\begin{array}{@{\,}ccccc@{\,}}
			 0 & 1     & 1     & 1     & 1\\
			 1 & 0     & j_3^2 & j_2^2 & 1\\
			 1 & j_3^2 & 0     & j_1^2 & 1\\
			 1 & j_2^2 & j_1^2 & 0     & 1\\
			 1 & 1     & 1     & 1     & 0
		\end{array}
		\right|.
\end{eqnarray}
\par
We now understand why this is a good approximation.
From (\ref{F}), we can observe that the leading contribution to $F(l_1,l_1)$ comes from the UV regions, i.e., $j_1,j_2,j_3,k_1,k_2,k_3 \sim l_1$.
In such a region, (\ref{WigEst}) is certainly valid.
\par
We can apply this approximation to $r\neq1$ cases as well.
The novelty here is the appearance of the ratio $r=l_2/l_1$:
\begin{eqnarray}
F(r)= 2 \int_0^2 dj_1 dj_2 dj_3 dk_1 dk_2 dk_3
\frac{2^6 \times j_1 j_2 j_3 k_1 k_2 k_3}
	{({j_1^2 \over r}+rk_2^2)({j_2^2 \over r}+rk_2^2)({j_3^2\over r}+rk_3^2)}
W(j_1,j_2,j_3)W(k_1,k_2,k_3) \label{FInt}.
\end{eqnarray}
$F(r)$ is symmetric under the exchange of $r$ with ${1\over r}$, which is consistent with the argument in the sub-section \ref{sec:tree1loop}.
\par
Numerically we can check whether this approximation is valid or not.
We perform the multiple integration (\ref{FInt}) twice analytically and eventually by Monte-Carlo integration\cite{MCInt1}\cite{MCInt2}.
The result is shown in fig.\ref{Fig:Fr}, which indicates this approximation is in good agreement with the method in the section \ref{subsec:intApprx}.
\par
We make a comment on the difference between the methods in the sections \ref{subsec:asymptotic} and \ref{subsec:intApprx}.
The former method requires more machine power with increasing $r$, while the latter does not\footnote{This causes discrepancy between the two methods around $r=9$ in the fig. \ref{Fig:EffactionVSr}.}.
So it seems that the latter is more economical and gives better results.
\par
It is noted that this approach cannot applies to the $S^2$ background case, that is, $r=0$ case. 
Setting $r=0(l_2=0)$, $F(l_1,l_2)$ becomes\cite{Imai:2003vr}
\begin{eqnarray}
F(l_1, l_2=0)=2({1\over l_1^3}) \sum_{j_1,j_2,j_3=1}^{2l_1}{1 \over l_1^3} \times
\frac{({2j_1\over l_1}+{1\over l_1})({2j_2\over l_1}+{1\over l_1})({2j_3\over l_1}+{1\over l_1})}
	{{j_1\over l_1}({j_1\over l_1}+{1\over l_1}){j_2\over l_1}({j_2\over l_1}+{1\over l_1}){j_3\over l_1}({j_3\over l_1}+{1\over l_1})}
\times l_1^3
	\left\{
		\begin{array}{ccc}
		j_1 & j_2 & j_3 \\
		l_1 & l_1 & l_1
		\end{array}
	\right\}^2.
\end{eqnarray}
Rescaling variables $j_{1,2,3}$ and taking the large $l_1$ limit, we have
\begin{eqnarray}
F(r=0)=2\left(\lim_{l_1 \to \infty} {1 \over l_1^3}\right)
	\int_0^2 dj_1 dj_2 dj_3{2^3 \over j_1 j_2 j_3} W(j_1,j_2,j_3). \label{F2D}
\end{eqnarray}
The factor $\left(\lim_{l_1 \to \infty} {1 \over l_1^3}\right)$ vanishes, while the integration part diverges because of the IR contributions, i.e. $j_1, j_2,j_3 \sim 0$.
This implies that (\ref{F2D}) is ill-defined, so that the naive approximation is not valid for the case $r=0$.
\par
This limitation can be explained from another point of view.
In the $r=0(l_2=0)$ case dominant contributions to $F(l_1,l_2)$ comes from IR regions, which is to say, $j_1, j_2, j_3 \sim 1$. 
In such a region  (\ref{WigEst}) is no longer valid.
This approximation cannot be applied to $S^2$ background.
In this case, however,  6j symbols are well-approximated by 3j symbols
\begin{eqnarray}
	\left\{
		\begin{array}{ccc}
		j_1 & j_2 & j_3 \\
		l_1 & l_1 & l_1
		\end{array}
	\right\}
\simeq
	{1\over \sqrt{2l}}
	\left(
		\begin{array}{ccc}
		j_1 & j_2 & j_3 \\
		0   & 0   & 0
		\end{array}
	\right).
\end{eqnarray}
\subsection{Higher order corrections \& total effective action}

In this sub-section, we determine the structure of the effective action to all orders.
By naive power counting we estimate the $i$-th loop contribution to be $O(n^{i+1} l_1^2 l_2^2 /(f^4 l_1 l_2)^{i-1})$ since we obtain a factor of $1/(l_1 l_2)$ from 6j symbols in the interaction vertices(see \cite{Imai:2003jb}).
Here we also assume that the $S^2 \times S^2$ background breaks SUSY softly, and so that leading corrections of each $i$-th loop cancel.
Under this assumption it appears that the $i$-th loop contribution becomes  $O(n^{i+1}l_1 l_2/(f^4 l_1 l_2)^{i-1})$.
In fact this order estimation is correct in the two loop corrections (\ref{2loop}).
This argument allows us to define the 't Hooft coupling of $U(n)$ gauge theory as $\lambda^2 \equiv (4 \pi)^2 n^2/(f^4 N)$ to all orders and so that the $i-$th loop contributions are $O(nN(\lambda^2)^{i-1})$.
\par
The above arguments give the effective action to all orders in terms of the 't Hooft coupling
\begin{eqnarray}
\Gamma(r)&=&\left[{(2 \pi)^2 \over \lambda^2}\left({{r + {1 \over r}} \over 2}\right)  
	+ {\tilde{h}}_4(r, \lambda^2) \right] n N, \label{effaction}\\
&& \mbox{with} \quad {\tilde{h}}_4(r, \lambda^2) =
	{F(r) \over 2}{ \lambda^2 \over (2 \pi)^2} + O(\lambda^4).
\end{eqnarray}
\par
Our remaining task is to minimize the effective action in terms of $n, r$, and $f$ or equivalently $\lambda^2$.
We can fix the 't Hooft coupling $\lambda$ by minimizing the effective action (\ref{effaction}) independently of $n$ (but dependent on $r$).
By solving $\partial \Gamma / \partial (\lambda^2) =0$ for $\lambda^2$ we find, up to two loop level,
\begin{eqnarray}
{{\bar{\lambda}}^2 \over (2 \pi)^2} = \sqrt{{1 \over F(r)}(r+{1 \over r})}.
\end{eqnarray}
\par
\begin{figure}
 \begin{center}
  \includegraphics[height=8cm]{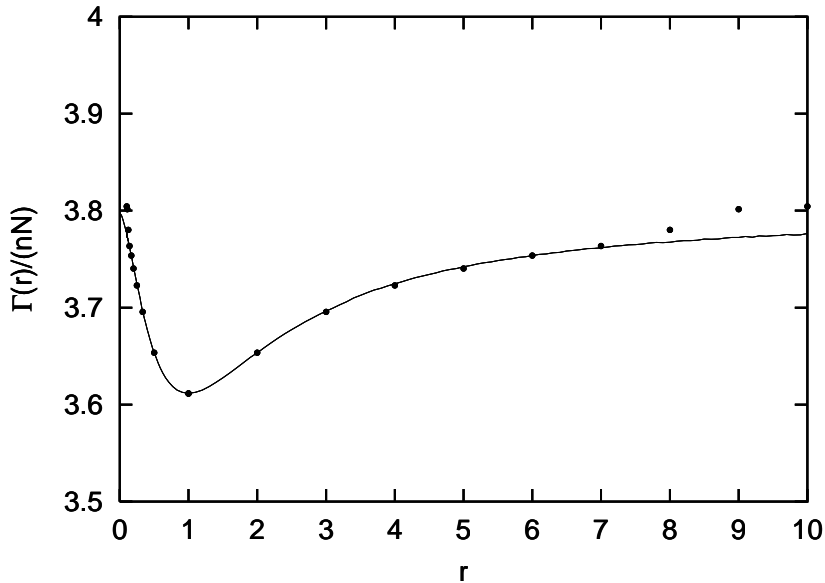}
 \end{center}
\caption{The effective action is plotted against $r(=l_2/l_1)$:The solid circles are obtained by the method described in the section \ref{subsec:asymptotic}, while the solid line is obtained by the method of the section \ref{subsec:intApprx}.}
\label{Fig:EffactionVSr}
\end{figure}

Then the effective action becomes
\begin{eqnarray}
\Gamma(r)\left.\right|_{\lambda=\bar{\lambda}} \simeq \sqrt{(r+{1 \over r})F(r)} n N. \label{Effaction}
\end{eqnarray}
This can be evaluated  numerically in two ways, i.e., the methods in the section \ref{subsec:asymptotic} and \ref{subsec:intApprx} (fig.\ref{Fig:EffactionVSr}).
It is found from this figure that the $r=1$ configuration minimizes the effective action, which means that the  configuration of the same size $S^2\times S^2$ is favored over the different size configurations.
Also it is apparent from (\ref{Effaction}) that the case $n=1$ is favored.
It is the most important result of this paper that the same size $S^2 \times S^2$ configuration with the $U(n=1)$ gauge group is singled out by the IIB matrix model.
It is worth noting that the effective action blows up, as we decrease $r$ below $r=1$.
This tendency is consistent with the result in \cite{Imai:2003jb}, that is, the configuration of the 4d limit($r=1$)  is favored rather than that of the 2d limit($r=0$).
\par
We expect that higher order corrections do not change this result by the argument in \cite{Imai:2003jb}.
\section{Conclusions and discussions}
In this paper we have investigated the effective action of the IIB matrix model on the fuzzy $S^2 \times S^2$ background up to two loop level.
In particular we have studied the case in which the size of one of $S^2 \times S^2$ is different from that of the other.
By the power counting and SUSY cancellation arguments, we can identify the 't Hooft coupling and the large $N$ scaling behavior of the effective actions to all orders.
In the large $N$ limit quantum corrections of the effective action are found to be $O(N)$.
This analysis will be applied to all order corrections of the effective action.
We have found the minimum point of the effective action at the two loop level,
and concluded that the same size
$S^2 \times S^2$ with $U(1)$ gauge group is favored.
\par
In the previous paper\cite{Imai:2003jb} we have concluded that the IIB matrix model favors the 4 dimensional spacetime(fuzzy $S^2 \times S^2$) rather than 2 and 6 dimensional spacetimes $S^2$ and $S^2 \times S^2 \times S^2$.
Combining the previous work and the results in this paper suggests that the IIB matrix model singles out not only {\it four dimensionality of spacetime} but also {\it more symmetric spacetime}.
We hope this suggestion is one of general features of the IIB matrix model.
\par
In addition we have developed a approach to calculate the amplitudes on the fuzzy $S^2 \times S^2$.
This approach cannot be applied to the $S^2$ background($r=0$) case.
We wish that extensions of this approach enables us to calculate amplitudes on other homogeneous spaces, such as $CP^2$ and $CP^3$\cite{MtxHom}.
\begin{center} \begin{large}
Acknowledgments
\end{large} \end{center}
We would like to thank Y. Kitazawa, D. Tomino for discussions.
This work is supported in part by the Grant-in-Aid for Scientific
Research from the Ministry of Education, Science and Culture of Japan.
\end{document}